\documentclass[%
 reprint,
 amsmath,amssymb,
 aps,
]{revtex4-2}

\usepackage[dvipdfmx]{graphicx}
\usepackage{dcolumn}
\usepackage{bm}
\usepackage{braket}
\usepackage{xcolor}

\begin{document}

\preprint{APS/123-QED}

\title{Quantum annealing in capacitively coupled Kerr parametric oscillators using frequency-chirped drives}

\author{T.~Yamaji$^{1, 2}$}
\thanks{Corresponding author: tyamaji@nec.com}
\author{S.~Masuda$^{2, 3}$}
\author{Y.~Kano$^{1, 2}$}
\author{Y.~Kawakami$^{1, 2}$}
\author{A.~Yamaguchi$^{1, 2}$}
\author{T.~Satoh$^{3}$}
\author{A.~Morioka$^{1, 2}$}
\author{Y.~Igarashi$^{1, 2}$}
\author{M.~Shirane$^{1, 2}$}
\author{T.~Yamamoto$^{1, 3}$}

\affiliation{$^1$Secure System Platform Research Laboratories, NEC Corporation, Kawasaki, Kanagawa 211-0011, Japan}
\affiliation{$^2$NEC-AIST Quantum Technology Cooperative Research Laboratory, National Institute of Advanced Industrial Science and Technology (AIST), Tsukuba, Ibaraki 305-8568, Japan}
\affiliation{$^3$Global Research and Development Center for Business by Quantum-AI technology (G-QuAT), National Institute of Advanced Industrial Science and Technology (AIST), Tsukuba, Ibaraki, Japan}

\date{\today}

\begin{abstract}
We study parametric oscillations of two capacitively coupled Kerr parametric oscillators (KPOs) with frequency-chirped two- and one-photon drives. 
The two-KPO system adiabatically evolves from the initial vacuum state to an oscillation state corresponding to a solution state in quantum-annealing applications. 
Frequency chirping dynamically changes the detuning between resonance and oscillation frequencies during parametric modulation and reduces unwanted population transfer to excited states caused by pure dephasing and photon loss. 
We observe that frequency chirping increases the success probability to obtain the solution state and that simulations taking into account pure dephasing reproduce experiments with and without frequency chirping. 
This study demonstrates the effectiveness and applicability of frequency chirping to a KPO-based quantum annealer.
\end{abstract}

\maketitle

\section{Introduction}
A Kerr parametric oscillator (KPO), a parametric oscillator in the single-photon Kerr regime \cite{Kirchmair2013}, is a promising candidate as the building block of quantum-information technologies such as gate-based quantum computation \cite{Goto2016a, Puri2017, Puri2020, Kanao2022, Xu2022, Masuda2022, Chono2022} and quantum annealing \cite{Goto2016, Nigg2017, Puri2017a, Goto2018, Zhao2018, Goto2019a, Onodera2020, Goto2020, Kewming2020, Kanao2021, Masuda2022}. 
Small-scale KPO systems have been developed to apply these technologies using Josephson parametric oscillators (JPOs) \cite{Wang2019, Grimm2020, Yamaji2022, Yamaji2023, Iyama2024, Hoshi2025, Frattini2024}.

\begin{figure}
	\includegraphics[width=\linewidth]{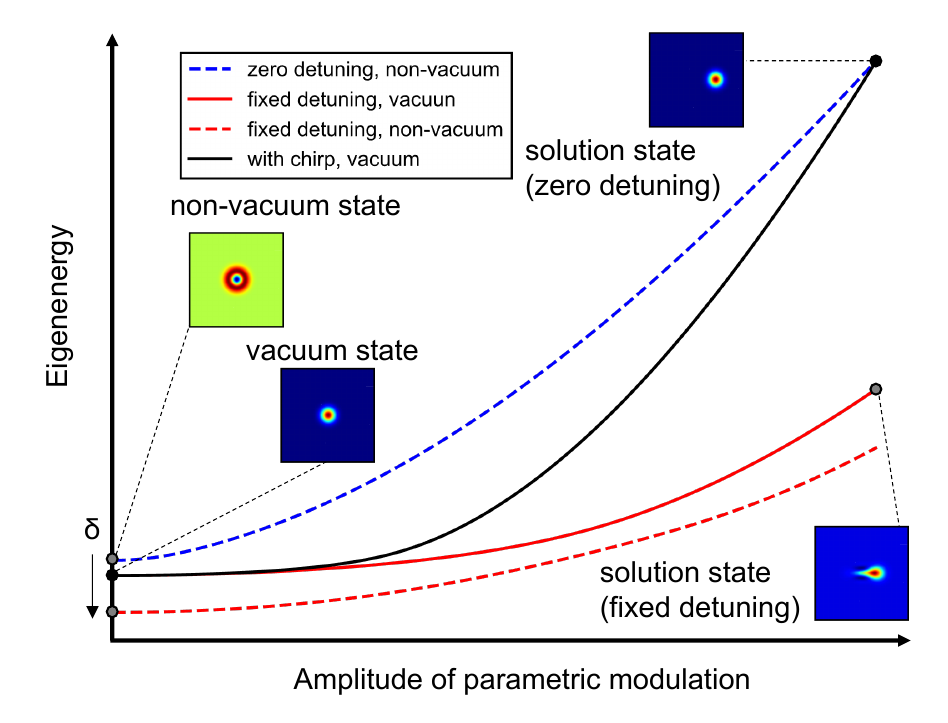}
	\caption{
Schematic energy level diagram of KPO-based quantum annealing.
We assume negative Kerr nonlinearity.
Solid and dashed lines show eigenenergies of states which adiabatically evolve from vacuum and non-vacuum states, respectively, as a function of the amplitude of parametric modulation.
Black, blue and red lines show eigenenergies with frequency chirping, zero detuning, and a fixed detuning $\delta<0$, respectively.
Insets show schematic one-bit Wigner functions at the start and the end of the parametric modulation.
}
	\label{fig:levels}
\end{figure}

Quantum annealing solves combinatorial optimization problems by searching for the global minima of Ising Hamiltonians \cite{Kadowaki1998, Farhi2000, Farhi2001, Albash2018}. 
The search is performed by gradually changing system parameters and adiabatically evolving the system from an initial state to the state corresponding to the solution, the solution state.
KPO-based quantum annealing uses the oscillation states of KPOs as Ising spins, and adiabatic evolution is achieved by applying parametric modulation, where the solution state corresponds to the ground state under parametric modulation in the frame rotating at the oscillation frequency.
Setting the initial state and tailoring the adiabatic evolution during parametric modulation appropriately can improve the success probability of KPO-based quantum annealing \cite{Goto2020, Kanao2021}.

The vacuum state is the most stable in the absence of parametric modulation, making it the natural choice as the initial state.
Searching for the solution via adiabatic transition requires that the vacuum state is the ground state under the initial condition, the initial ground state.
However, due to perturbations such as couplings between the KPOs, the vacuum state may not be the initial ground state.
Figure~\ref{fig:levels} shows the schematic energy level diagram of quantum annealing using KPOs in a frame rotating at the oscillation frequency. 
When a non-vacuum state is the initial ground state (the top left inset), the solution state cannot be obtained by adiabatic evolution from the vacuum state (the blue dashed line).
Sufficiently large detuning between resonance and oscillation frequencies shifts the eigenenergies of the non-vacuum states, making the vacuum state the initial ground state (the bottom left inset) \cite{Goto2018}. 
In this case, the vacuum state can adiabatically evolve to the solution state with increased amplitude of parametric modulation (the red lines) \cite{Goto2016a, Goto2016}.  
However, the large detuning has the drawback of shifting the solution state away from coherent states, which are resistant to one-photon loss (the bottom right inset).
Frequency chirping, which dynamically changes the detuning during parametric modulation, is expected to combine the advantages of both conditions with and without detuning (the black solid line) \cite{Puri2017a, Goto2020}. 
Namely, it enables adiabatic evolution from the vacuum state to the solution state close to coherent states (the top right inset).

In the field of superconducting quantum circuits, frequency chirping has been conducted using a capacitively shunted Josephson tunnel junction to study quantum fluctuations \cite{Murch2011}, and previous experiments used frequency chirping for KPO-transmon systems to offset alternating-current Stark shift in gate-based quantum-computation applications \cite{Iyama2024, Hoshi2025, Xu2025}. 
However, no frequency-chirping experiments have been conducted for KPO-based quantum-annealing applications.

Here we study simultaneous parametric oscillations of two capacitively coupled JPOs in the single-photon Kerr regime using frequency-chirped two- and one-photon drives for KPO-based quantum-annealing applications. 
The two-photon drives excite parametric oscillations, and the one-photon drives control {\it in situ} the occurrence probabilities of the self-oscillating states, called phase locking. 
The capacitive coupling correlates the oscillation phases of the KPOs, where the strength and sign of the correlation are tunable via the phase difference of the two-photon drives applied to the KPOs \cite{Masuda2022, Yamaji2023}. 
The coupled-KPO system behaves as an Ising machine, the bit-to-bit coupling and local fields of which are encoded via the two- and one-photon drives, respectively. 
During parametric modulation, frequency chirping is synchronously applied to the drives to shift the detuning from a negative value to zero. 
The frequency-chirped drives enhance the phase-locking probabilities of each KPO and the correlation between parametric oscillations compared with without applying frequency chirping, which leads to a higher success probability to obtain the correct solution of a two-spin Ising Hamiltonian. 
The experimental results are consistent with numerical simulations solving the master equation with pure dephasing taken into account. 
This study demonstrates the efficient manipulation of KPOs using frequency-chirped drives, which can be applied to a novel quantum-annealing system composed of a KPO network.

\section{Device}

\begin{figure}
	\includegraphics[width=\linewidth]{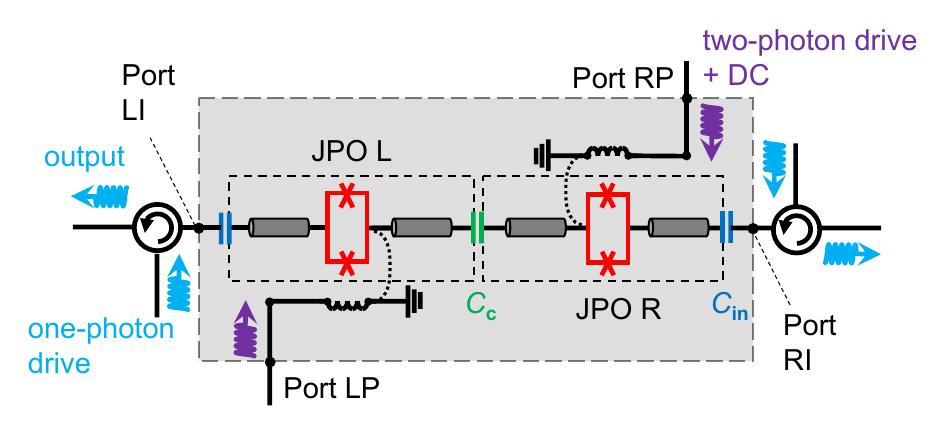}
	\caption{
Schematic circuit diagram of device chip. 
Grey rectangular region represents chip. 
Each JPO labeled L and R consists of symmetric DC-SQUID (superconducting quantum interference device) (shown in red) embedded in center of transmission-line resonator (grey tubes). 
Each DC-SQUID is inductively coupled to on-chip pump line. 
Coupling capacitor $C_{\rm c}$ and input capacitors $C_{\rm in}$ are shown in green and blue, respectively. 
Four ports of chip are labeled as LI, LP, RI, and RP. 
Circulators outside LI and RI ports route output microwaves. 
}
	\label{fig:sample}
\end{figure}
 
Figure~\ref{fig:sample} shows the circuit diagram of the device chip, which consists of two capacitively coupled JPOs on the left and right sides labeled L and R, respectively. 
The geometric design of each JPO is the same as in our previous experiments \cite{Yamaji2022, Yamaji2023} (see Appendix~\ref{sup:device_detail} for details). 
The direct current (DC) and two-photon drives with angular frequencies $\omega_{\rm p\scalebox{0.5}{L(R)}}$ are applied to the pump lines to tune the resonance frequencies $\omega_{\rm r\scalebox{0.5}{L(R)}}$ and excite parametric oscillations, respectively. 
The oscillation frequencies are equal to half the pump frequencies $\omega_{\rm p\scalebox{0.5}{L(R)}}/2$, and the one-photon drives at the oscillation frequencies are applied to the input capacitors to control the occurrence probabilities of the oscillation states. 
Hereafter, all the input/output powers of the JPOs are expressed by their values at the relevant ports on the chip.

The capacitive coupling correlates parametric oscillations of the JPOs under the condition that the pump frequencies are the same, $\omega_{\rm p\scalebox{0.5}{L}}=\omega_{\rm p\scalebox{0.5}{R}}\equiv \omega_{\rm p}$ ($\omega_{\rm p}$-resonance condition).
The oscillation state of each JPO is either one of the coherent states $\ket{\pm \alpha_{\scalebox{0.5}{L(R)}}}$, which have an equal amplitude $\alpha_{\scalebox{0.5}{L(R)}}$ and opposite phases, $0$ or $\pi$, relative to the two-photon drive \cite{Nayfeh1995, Dykman1998, Strogatz2000}. 
The simultaneous oscillation states of the JPOs can be written as a tensor product of the coherent states, $\ket{s_{\rm \scalebox{0.5}{L}}\alpha_{\rm \scalebox{0.5}{L}}}\ket{s_{\rm \scalebox{0.5}{R}}\alpha_{\rm \scalebox{0.5}{R}}}$, where tensor products are defined as $\ket{x}\ket{y} \equiv \ket{x}_{\rm \scalebox{0.5}{L}}\ket{y}_{\rm \scalebox{0.5}{R}}$, $s_j=\pm 1$ is the Ising spin corresponding to the oscillation state $\ket{\pm \alpha_j}$, and the index $j=$L/R represents the label of the JPOs. 
The eigenenergies of the simultaneous oscillation states have a term dependent on $s_j$ (see Appendix~\ref{sup:op.point} for details),
\begin{eqnarray}
- \left[-J_{\scalebox{0.5}{\rm LR}}s_{\rm \scalebox{0.5}{L}} s_{\rm \scalebox{0.5}{R}} + h_{\rm \scalebox{0.5}{L}}s_{\rm \scalebox{0.5}{L}} + h_{\rm \scalebox{0.5}{R}}s_{\rm \scalebox{0.5}{R}} \right], \label{eq:Ising}
\end{eqnarray}
where $J_{\scalebox{0.5}{\rm LR}}=2{\rm cos}\left(\theta_{\rm p}/2\right)\alpha_{\scalebox{0.5}{\rm L}}\alpha_{\scalebox{0.5}{\rm R}}g$ is the bit-to-bit coupling, $h_{j}=2{\rm sin}\theta_{{\rm s}j} \alpha_j \Omega_{{\rm d}j}$ is the local field, $\Omega_{{\rm d}j} \equiv \sqrt{P_{{\rm s}j} \kappa_{{\rm e}j}/(\hbar \omega_{{\rm r}j})}$ is the Rabi frequency of the one-photon drive, $P_{{\rm s}j}$ is the power of the one-photon drive (hereafter called signal power), $\kappa_{{\rm e}j}$ is the external photon-loss rate, $\theta_{{\rm s}j}$ is the phase of the one-photon drive relative to the two-photon drive (signal phase), $g$ is the coupling between the two JPOs, and $\theta_{\rm p}$ is the phase difference between the two-photon drives (pump phase). 
The term represents an Ising energy $E_{\rm Ising}=\left[-J_{\scalebox{0.5}{\rm LR}}s_{\rm \scalebox{0.5}{L}} s_{\rm \scalebox{0.5}{R}} + h_{\rm \scalebox{0.5}{L}}s_{\rm \scalebox{0.5}{L}} + h_{\rm \scalebox{0.5}{R}}s_{\rm \scalebox{0.5}{R}} \right]$, which is easily programmable via $\theta_{\rm p}$, $\theta_{{\rm s}j}$, and $\Omega_{{\rm d}j}$. The simultaneous oscillation state with the highest eigenenergy (the lowest Ising energy) is favored due to the negative Kerr nonlinearity and is referred to as the solution state of the encoded Ising Hamiltonian.

The initial state of the system, $\ket{0}\ket{0}$, where $\ket{0}$ is the vacuum state, evolves to the solution state via adiabatic transition or relaxation. 
The adiabatic transition requires the initial condition that the vacuum state is the highest energy state of the Hamiltonian without the drives, requiring $\Delta_j < -g < 0$, where $\Delta_j\equiv \omega_{{\rm r}j} -\omega_{\rm p}/2$ is the detuning of the resonance frequency from the oscillation frequency \cite{Goto2018}. When the initial condition is satisfied, the solution state can be found by gradually increasing the drive amplitudes from zero to values determined from the encoding of $J_{\scalebox{0.5}{\rm LR}}$ and $h_{j}$. 
When the initial condition is violated, the evolution to the solution state occurs only via slow relaxation to the steady state, which is mediated by processes such as pure dephasing and one-photon loss. The relaxation requires longer parametric modulation than the adiabatic transition, which decreases the probability to obtain the solution state due to the phase flip of parametric oscillations \cite{Mirrahimi2014, Puri2017}.

In the two-KPO experiments presented later, we set an operating point well in the single-photon Kerr regime using the $\omega_{\rm r}$-resonance condition $\omega_{{\rm r}\scalebox{0.5}{L}}=\omega_{{\rm r}\scalebox{0.5}{R}}=\omega_{\rm r}$ ($\Delta_{\rm \scalebox{0.5}{L}}=\Delta_{\rm \scalebox{0.5}{R}}\equiv \Delta$) for the simplicity of the measurement (see also Appendix~\ref{sup:op.point} for details).

\section{Phase locking}

\begin{figure}
	\includegraphics[width=1.0 \linewidth]{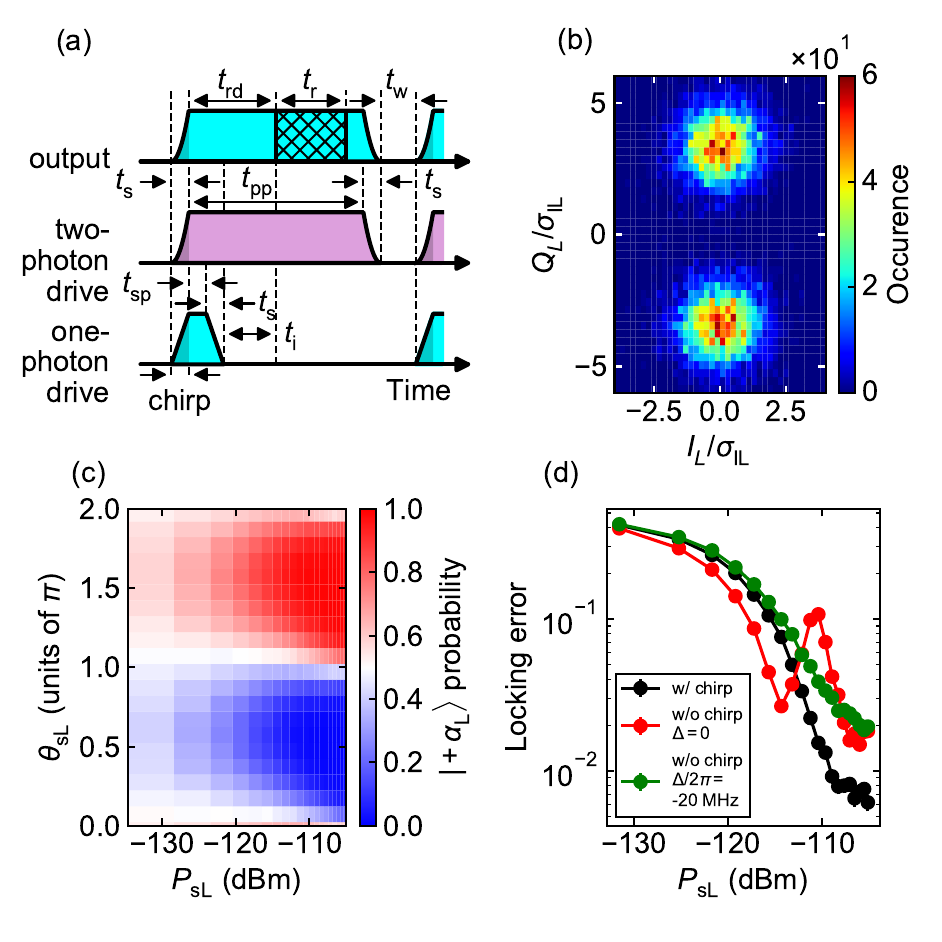}
	\caption{
Phase locking of JPO (KPO) L. 
(a) Pulse sequence of two-photon drive, one-photon drive, and output of parametric oscillations. 
Cyan and purple represent frequencies of pulses $\omega_{\rm p}/2$ and $\omega_{\rm p}$, respectively. 
Shaded and hatched regions represent frequency chirping and integration time of heterodyne measurement, respectively (see Appendix~\ref{sup:pulse_parameter} for details). 
(b) $IQ$-plane histogram of output with frequency chirping. 
Relative phases of $IQ$ amplitudes are subtracted analytically to align two peaks on $Q$ axis. 
Magnitudes of $IQ$ amplitudes are normalized by standard deviation of $I$ amplitudes. 
(c) $P_{\rm s \scalebox{0.5}{L}}$ and $\theta_{\rm s \scalebox{0.5}{L}}$ dependence of occurrence probabilities of $\ket{+\alpha_{\rm \scalebox{0.5}{L}}}$ with frequency chirping. 
(d) $P_{\rm s \scalebox{0.5}{L}}$ dependence of locking error. 
Locking errors are deduced by subtracting maximum observed probability of $\ket{+\alpha_{\rm \scalebox{0.5}{L}}}$ at each $P_{\rm s\scalebox{0.5}{L}}$ from unity. 
Black data shows locking error with frequency chirping, which is deduced from data shown in (c). 
Red and green data show locking errors deduced from experiments (not shown) using same experimental parameters as (c) without frequency chirping at $\Delta=0$ and $\Delta/2\pi=-20$~MHz, respectively. 
	}
	\label{fig:phaselock}
\end{figure}

We conducted a one-KPO experiment with JPO L to study the effect of frequency chirping on phase locking using the pulse sequence shown in Fig.~\ref{fig:phaselock}(a).
The trapezoidal two- and one-photon drives are applied simultaneously, and then the output is measured using the heterodyne measurement. 
The detuning $\Delta/2\pi$ shifts linearly from $\Delta_0/2\pi =-20$~MHz to zero during the rising slopes of the trapezoidal pulses by synchronously chirping the frequencies of the two- and one-photon drives while maintaining the relationship that the two-photon-drive frequency is twice the one-photon-drive one (see Appendix~\ref{sup:pulse_parameter} for the details of the pulse sequence). 
The operating point of the experiment was the same as the following two-KPO experiments except that JPO R was far detuned and not pumped. 

Figure~\ref{fig:phaselock}(b) shows the histogram of the in-phase and quadrature ($IQ)$ amplitudes obtained from the heterodyne measurement under the condition that the occurrence probabilities of the oscillation states $\ket{\pm\alpha_{\rm \scalebox{0.5}{L}}}$ are almost the same. 
The two peaks with the same amplitude and an opposite phase correspond to $\ket{\pm\alpha_{\rm \scalebox{0.5}{L}}}$. 
Figure~\ref{fig:phaselock}(c) shows the $P_{\rm s \scalebox{0.5}{L}}$ and $\theta_{\rm s \scalebox{0.5}{L}}$ dependence of the observed probabilities of $\ket{+\alpha_{\rm \scalebox{0.5}{L}}}$ with frequency chirping, where the maximum $P_{\rm s \scalebox{0.5}{L}}$, $-105$~dBm, corresponds to the local field $h_{\rm \scalebox{0.5}{L}}$ with the same order as the coupling $J_{\rm \scalebox{0.5}{LR}}$.  
As shown in Eq.~(\ref{eq:Ising}), the KPO is locked to $\ket{+(-)\alpha_{\rm \scalebox{0.5}{L}}}$ at $\theta_{\rm s \scalebox{0.5}{L}}=3\pi/2(\pi/2)$ (phase locking). 
The probability to obtain the state favored by phase locking, referred to as the locking probability, is an increasing function of $P_{\rm s \scalebox{0.5}{L}}$. 
We also refer to the remainder probability after subtracting the locking probability from unity as the locking error. 
Figure~\ref{fig:phaselock}(d) compares the $P_{\rm s \scalebox{0.5}{L}}$ dependence of the locking error with and without frequency chirping. 
Frequency chirping reduces the locking error in the high-signal-power region. 
The locking error with frequency chirping at $P_{\rm s \scalebox{0.5}{L}}=-105$~dBm is $0.62 \pm 0.08\%$, where the uncertainty represents the statistical uncertainty (standard deviation). 
We attribute the locking error to pure dephasing and one-photon loss during parametric modulation. 
The locking error with frequency chirping is also systematically smaller than that without frequency chirping at $\Delta=-20$~MHz in the low-signal-power region. 
Frequency chirping can increase the probability to obtain the solution state for practical problem Hamiltonians with local fields of moderate amplitude. 
It is worth noting that the locking error without frequency chirping at $\Delta=0$ is not a monotonically decreasing function of the signal power because the initial vacuum state cannot adiabatically evolve to the solution state.

\begin{figure}
	\includegraphics[width=1.0 \linewidth]{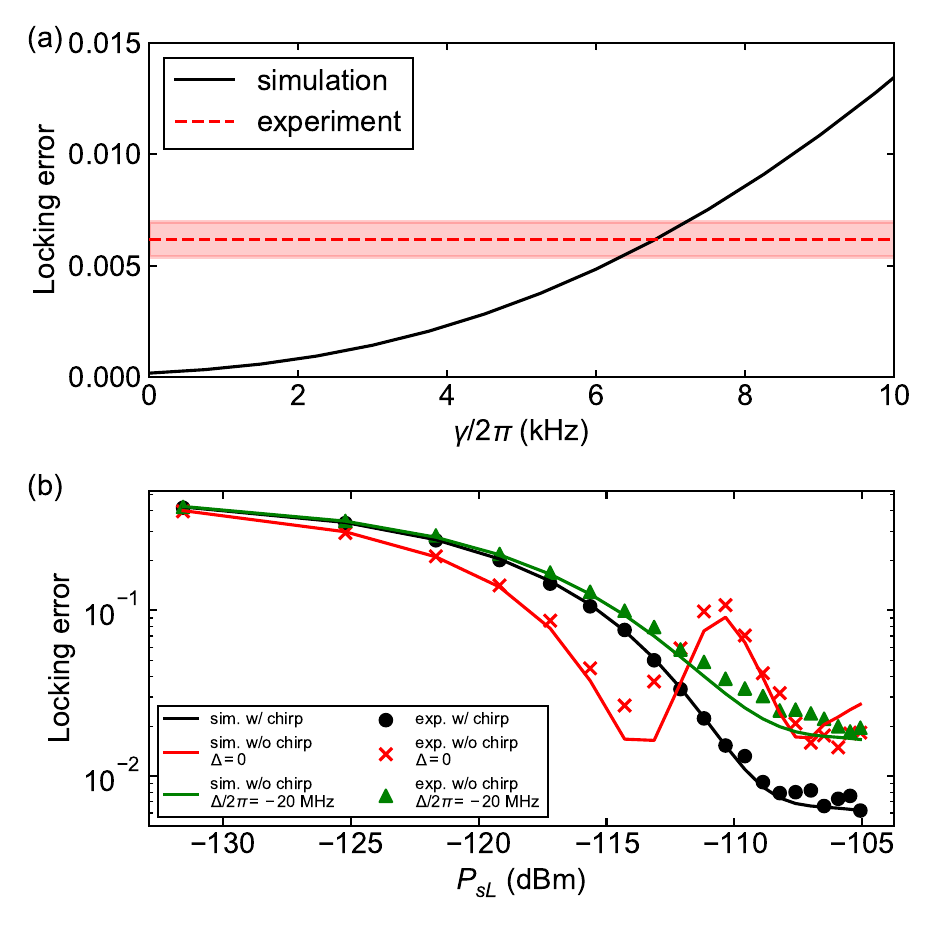}
	\caption{
Simulation of phase locking. 
(a) Simulated pure-dephasing-rate dependence of locking error with frequency chirping. 
Simulation parameters are fixed to experimental ones used for data with frequency chirping and highest signal power shown in Fig.~\ref{fig:phaselock}(d), where we fix $\theta_{\rm s\scalebox{0.5}{L}}=3\pi/2$. 
Dashed red line with band represents corresponding experimental locking error with statistical uncertainty (standard deviation). 
(b) Simulated signal-power dependence of locking error at $\gamma_{\rm \scalebox{0.5}{L}}/2\pi=6.8$~kHz. 
Markers and solid lines show experimental data shown in Fig.~\ref{fig:phaselock}(d) and simulated data, respectively. 
Simulation parameters are fixed to experimental ones used for data in Fig.~\ref{fig:phaselock}(c), and simulated locking errors are deduced in same manner as in Fig.~\ref{fig:phaselock}(d). 
Black circles show locking error with frequency chirping. 
Red crosses and green triangles show locking errors obtained without frequency chirping at $\Delta=0$ and $\Delta/2\pi=-20$~MHz, respectively. 
	}
	\label{fig:1bit_simulation}
\end{figure}

The experimental data are compared with the simulation, as shown in Fig.~\ref{fig:1bit_simulation}. 
The simulation solves the Lindblad master equation with the one-JPO Hamiltonian shown in Eq.~(\ref{eq:H_one}) and fixes all the parameters to the experimental ones, except the pure dephasing rate $\gamma_{\rm \scalebox{0.5}{L}}$, which is not experimentally measured (see Appendix~\ref{sup:numerical} for details). 
As shown in Fig.~\ref{fig:1bit_simulation}(a), the locking error is an increasing function of $\gamma_{\rm \scalebox{0.5}{L}}$ because pure dephasing causes the phase flip of parametric oscillations \cite{Mirrahimi2014, Puri2017}. 
The minimum locking error obtained in the experiment corresponds to $\gamma_{\rm \scalebox{0.5}{L}}/2\pi=6.8\pm 0.5$~kHz, where the uncertainty comes from the statistical uncertainty. 
The probability that the KPO is not initialized to the vacuum state at the start of the pulse sequence, referred to as the initialization error, is estimated to be 0.1\%, which is determined by the duration of the pulse-sequence interval.
We neglected the initialization error caused by noise photons because we did not observe transitions from excited states in spectroscopy measurements using a weak excitation power. 
The locking error caused by the initialization error is negligible because the uninitialized state, mainly the one-photon Fock state $\ket{1}$, can evolve to the locked state via relaxation with a simulated probability of 97\% at the signal power of -105~dBm. 
Figure~\ref{fig:1bit_simulation}(b) shows the comparison between the experimental data in Fig.~\ref{fig:phaselock}(d) and simulation with $\gamma_{\rm \scalebox{0.5}{L}}/2\pi=6.8$~kHz. 
The simulation qualitatively reproduces the experimental data with and without frequency chirping.

\section{Simultaneous parametric oscillations}

\begin{figure*}
	\includegraphics[width=1.0 \linewidth]{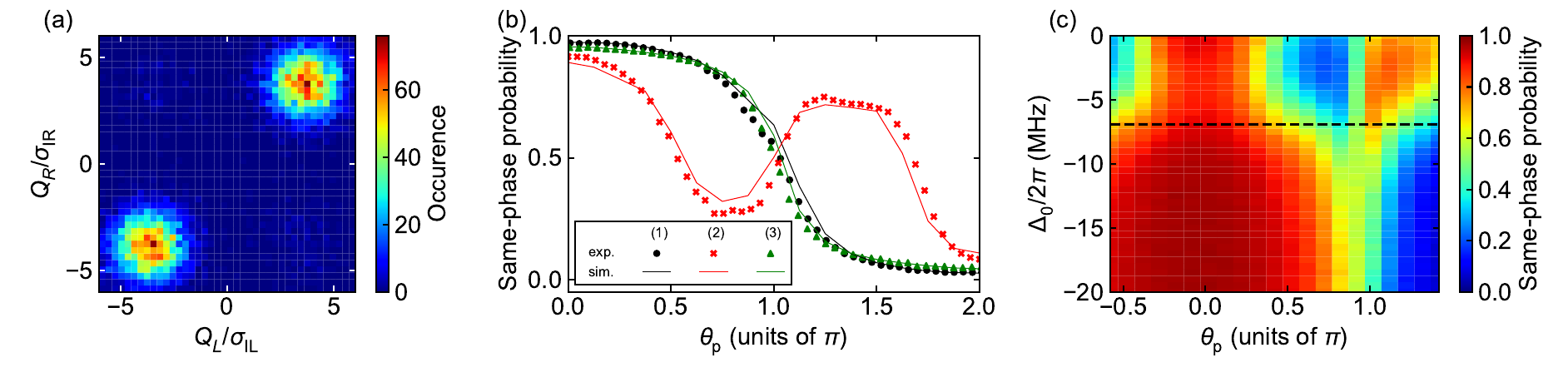}
	\caption{
Simultaneous parametric oscillations without one-photon drives. 
Trapezoidal two-photon drives are simultaneously applied to KPOs, then outputs are measured  (see Appendix~\ref{sup:pulse_parameter} for details). 
(a) Histogram of $Q$ amplitudes of outputs with frequency chirping. 
$Q$ amplitude of each KPO is normalized by standard deviation of $I$ amplitude. 
(b) Pump phase $\theta_{\rm p}$ dependence of same-phase probability. 
Same-phase probability is sum of observed probabilities of same-phase states $\ket{\pm \alpha_{\rm \scalebox{0.5}{L}}}\ket{\pm\alpha_{\rm \scalebox{0.5}{R}}}$. 
Black circles (1) show experimental data obtained with frequency chirping. 
Red crosses and green triangles (2, 3) show experimental data obtained without frequency chirping at $\Delta=0$ and $\Delta/2\pi=-20$~MHz, respectively. 
Solid lines show simulated dependence with and without frequency chirping, where we assume $\gamma_{\rm \scalebox{0.5}{L(R)}}/2\pi=7.7$~kHz. 
(c) Detuning dependence of same-phase probability with frequency chirping. 
Horizontal dashed line shows coupling strength $-g$. 
	 }
	\label{fig:correlation}
\end{figure*}

Next, we conducted the two-KPO experiment without one-photon drives to study the effect of frequency chirping on the correlation between simultaneous parametric oscillations. 
The simultaneous parametric oscillations without one-photon drives are described with the Ising Hamiltonian Eq.~(\ref{eq:Ising}) without local fields. 
Figure~\ref{fig:correlation}(a) shows the histogram of the $Q$ amplitudes of both KPOs for $\theta_{\rm p}=0.0\pi$. 
The oscillation states with the same phases, $\ket{\pm \alpha_{\rm \scalebox{0.5}{L}}}\ket{\pm\alpha_{\rm \scalebox{0.5}{R}}}$ (the first and third quadrants), occur with higher probabilities than those with different phases, $\ket{\pm \alpha_{\rm \scalebox{0.5}{L}}}\ket{\mp\alpha_{\rm \scalebox{0.5}{R}}}$ (the second and fourth quadrants). 
The same-phase correlation originates from the capacitive coupling between the KPOs, whose bit-to-bit coupling is ferromagnetic $J_{\rm \scalebox{0.5}{LR}}>0$ at $\theta_{\rm p}=0.0\pi$. 
The occurrence probabilities of $\ket{+ \alpha_{\rm \scalebox{0.5}{L}}}\ket{+\alpha_{\rm \scalebox{0.5}{R}}}$ and $\ket{- \alpha_{\rm \scalebox{0.5}{L}}}\ket{-\alpha_{\rm \scalebox{0.5}{R}}}$ are the same when one-photon drives are absent.

Figure~\ref{fig:correlation}(b) shows the pump phase $\theta_{\rm p}$ dependence of the observed probability of the same-phase states, referred to as the same-phase probability, with and without frequency chirping. 
The cosine-like $\theta_{\rm p}$ dependence with frequency chirping shows the tunability of the bit-to-bit coupling, $J_{\rm \scalebox{0.5}{LR}}\propto {\rm cos}\left(\theta_{\rm p}/2\right)$ \cite{Yamaji2023}. 
The maximum correlation (same-phase or opposite-phase) with frequency chirping is $97.3 \pm 0.2\%$, where the uncertainty represents the statistical uncertainty (standard deviation). 
The dependence obtained without frequency chirping at $\Delta=0$ and $\Delta/2\pi=-20$~MHz shows a correlation smaller than that with frequency chirping (91.6 and 95.5\%, respectively). 
The $\theta_{\rm p}$ dependence without frequency chirping at $\Delta=0$ is not cosine shaped because the initial vacuum state is not the highest energy state, and the adiabatic transition from the initial vacuum state to correlated oscillation states is impossible. 
It is worth noting that the $\theta_{\rm p}$ dependence without frequency chirping at $\Delta=0$ becomes cosine shaped when the pulse sequence is much longer, as shown in our previous experiment \cite{Yamaji2023}, because the two-KPO system completely relaxes to the steady state.

Figure~\ref{fig:correlation}(c) shows the experimental detuning dependence of the same-phase probability, where $\theta_{\rm p}$ and the detuning at the start of frequency chirping $\Delta_0$ are swept. 
When the detuning at the start satisfies $\Delta_0 < -g$, the vacuum state $\ket{0}\ket{0}$ is the highest energy state, and the adiabatic evolution from the vacuum state to correlated states is possible. 
The measured $\theta_{\rm p}$ dependence is cosine-like when $\Delta_0$ is sufficiently smaller than $-g$ and not when $\Delta_0 > -g$. 
Although in this study we fixed the qualitative profile of frequency chirping, linear in time, frequency chirping may be further optimized using more complex profiles, such as nonlinear ones, which is an issue for future studies.

We compared the experimental $\theta_{\rm p}$ dependence with the simulation solving the Lindblad master equation. 
In the same manner as the simulation for the one-KPO experiment, the simulation uses the experimentally obtained parameters, and there are no free parameters except the pure dephasing rate $\gamma_j$ (see Appendix~\ref{sup:numerical} for details). 
We assume in the simulation that the KPOs have the same dephasing rate since the design and DC-flux-bias gradient of $\omega_{{\rm r}j}$ of the JPOs are almost the same (See also Appendices A and B). 
The pure dephasing rate is estimated to be $\gamma_{\rm \scalebox{0.5}{L(R)}}/2\pi=7.7\pm 0.6$~kHz from the experimental maximum correlation obtained using frequency chirping. 
The estimation of $\gamma_{\rm \scalebox{0.5}{L(R)}}$ has systematic uncertainty due to the initialization error, which is estimated to be 2.2\% from the simulation. 
Although the uninitialized states (mainly $\ket{0}\ket{1}$ and $\ket{1}\ket{0}$ states) also evolve to the correlated states with a probability of 93\% via relaxation, the initialized error reduces the maximum correlation with frequency chirping by 0.2\% and enhances the estimation of $\gamma_{\rm \scalebox{0.5}{L(R)}}/2\pi$ by 0.6~kHz. 
The estimated pure dephasing rate is consistent with the one-KPO experiment within the statistical and systematic uncertainties.  
The simulations with $\gamma_{\rm \scalebox{0.5}{L(R)}}/2\pi=7.7$~kHz are qualitatively consistent with the experiment. 
The estimated $\gamma_j$ is five times smaller than in our previous experiment \cite{Yamaji2023}. 
The reason is currently not clear, and experimental investigation of pure dephasing using time-domain measurements is also required for future studies \cite{Abdumalikov2011, Lu2021}.

\begin{figure}
	\includegraphics[width=1.0 \linewidth]{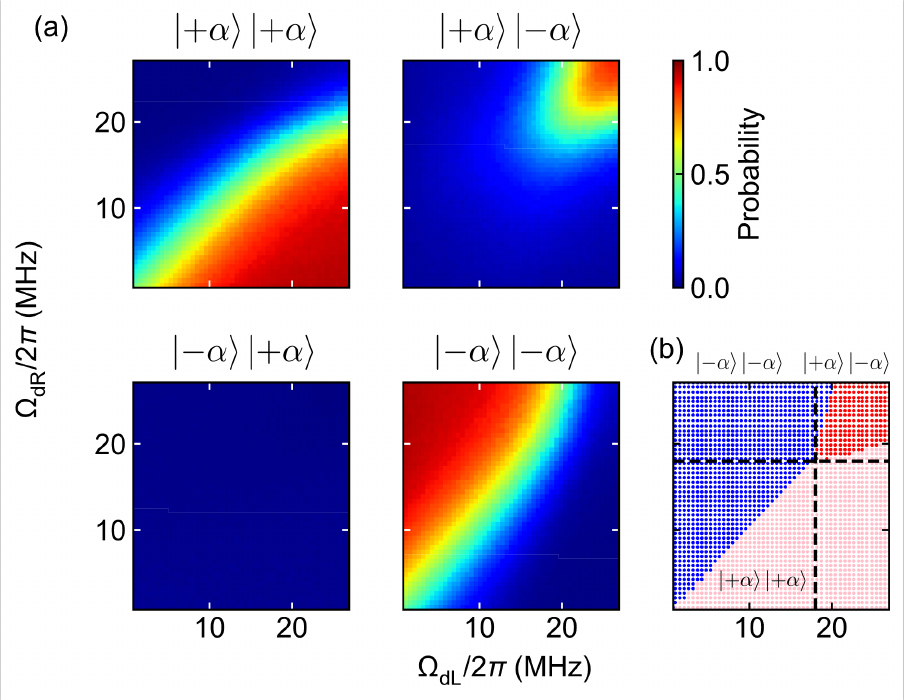}
	\caption{
Simultaneous parametric oscillations with one-photon drives and frequency chirping, where $J_{\rm \scalebox{0.5}{LR}}>0$ and $h_{\rm \scalebox{0.5}{L}} < 0 < h_{\rm \scalebox{0.5}{R}}$.  
Trapezoidal two- and one-photon drives are applied simultaneously to KPOs, and outputs from KPOs are measured (see Appendix~\ref{sup:pulse_parameter} for details).
(a) Rabi-frequency dependence of observed probabilities of simultaneous oscillation states. 
Color scale shared among all plots shows observed probabilities of state indicated at top of each plot. 
(b) Distribution of oscillation state with highest observed probability. 
Pink, red, and blue circles indicate parameter region where $\ket{+ \alpha_{\rm \scalebox{0.5}{L}}}\ket{+\alpha_{\rm \scalebox{0.5}{R}}}$, $\ket{+ \alpha_{\rm \scalebox{0.5}{L}}}\ket{-\alpha_{\rm \scalebox{0.5}{R}}}$, and $\ket{- \alpha_{\rm \scalebox{0.5}{L}}}\ket{-\alpha_{\rm \scalebox{0.5}{R}}}$ states respectively have highest probability. 
Dashed line shows $\Omega_{\rm d\scalebox{0.5}{L(R)}}/2\pi=18$~MHz. 
}
	\label{fig:boltzmann}
\end{figure}

The one- and two-KPO experiments above correspond to Ising models with a local field and bit-to-bit coupling, respectively. 
We also conducted an experiment combining these experiments, the two-KPO experiment with one-photon drives, to demonstrate quantum annealing with a general two-spin Ising model using frequency chirping. 
Figure~\ref{fig:boltzmann} shows the probability distribution of the oscillation states with one-photon drives as a function of the amplitudes of these drives (the Rabi frequencies). 
We set the phases of the two- and one-photon drives to $\theta_{\rm p}=0.0\pi$ and $\theta_{\rm s\scalebox{0.5}{L(R)}}=1.5\pi (0.5\pi)$, respectively, where bit-to-bit coupling and local fields reduce the Ising energies of the same-phase states $\ket{\pm \alpha_{\rm \scalebox{0.5}{L}}}\ket{\pm \alpha_{\rm \scalebox{0.5}{R}}}$ and opposite-phase state $\ket{+ \alpha_{\rm \scalebox{0.5}{L}}}\ket{-\alpha_{\rm \scalebox{0.5}{R}}}$, respectively. 
Since the states favored by bit-to-bit coupling and local fields differ, the oscillation state with the highest occurrence probability depends on the relative magnitudes of bit-to-bit coupling and local fields, as shown in Fig.~\ref{fig:boltzmann}(a). 
When both local fields are larger than bit-to-bit coupling ($\left|h_{j}\right| > J_{\scalebox{0.5}{LR}}$), the $\ket{+ \alpha_{\rm \scalebox{0.5}{L}}}\ket{-\alpha_{\rm \scalebox{0.5}{R}}}$ state has the highest probability. 
However, the $\ket{+ \alpha_{\rm \scalebox{0.5}{L}}}\ket{+ \alpha_{\rm \scalebox{0.5}{R}}}$ state has the highest probability under the conditions $\left|h_{\rm \scalebox{0.5}{R}}\right| < J_{\scalebox{0.5}{LR}}$ and $\left|h_{\rm \scalebox{0.5}{R}}\right| < \left|h_{\rm \scalebox{0.5}{L}}\right|$. 
Likewise, the $\ket{- \alpha_{\rm \scalebox{0.5}{L}}}\ket{- \alpha_{\rm \scalebox{0.5}{R}}}$ state has the highest probability under the conditions $\left|h_{\rm \scalebox{0.5}{L}}\right| < J_{\scalebox{0.5}{LR}}$ and $\left|h_{\rm \scalebox{0.5}{L}}\right| < \left|h_{\rm \scalebox{0.5}{R}}\right|$. 
The $\ket{- \alpha_{\rm \scalebox{0.5}{L}}}\ket{+\alpha_{\rm \scalebox{0.5}{R}}}$ state always has the lowest probability since both bit-to-bit coupling and local fields disfavor the state. 
Figure~\ref{fig:boltzmann}(b) shows the state with the highest observed probability as a function of the Rabi frequencies. 
The state with the highest observed probability changes at $\Omega_{{\rm d}j}/2\pi=18\ {\rm MHz}$. 
The simulation involving $\gamma_{\rm \scalebox{0.5}{L(R)}}/2\pi=7.7$~kHz and $\Omega_{\rm d \scalebox{0.5}{L}} = \Omega_{\rm d \scalebox{0.5}{R}}$ also shows that the $\ket{+ \alpha_{\rm \scalebox{0.5}{L}}}\ket{-\alpha_{\rm \scalebox{0.5}{R}}}$ state has the highest probability at $\Omega_{{\rm d}j}/2\pi>18\ {\rm MHz}$. 
The measurement demonstrates quantum annealing using the two-KPO system with ferromagnetic bit-to-bit coupling and controllable local fields.

\section{Conclusion}
We studied the system composed of two capacitively coupled JPOs in the single-photon Kerr regime using frequency-chirped two- and one-photon drives. 
We observed that frequency chirping enhances the locking probabilities and correlation between simultaneous parametric oscillations, which correspond to the enhancement of the probability to obtain the correct solution of an Ising Hamiltonian with a local field and bit-to-bit coupling, respectively. 
The enhancement is because frequency chirping reduces population transfer from the ground state to excited state during parametric modulation. 
The locking error was less than 1\% at a signal power of -105~dBm, and the correlation of 97\% was achieved using frequency chirping, which is higher than in our previous experiments \cite{Yamaji2022, Yamaji2023}. 
The experimental data are consistent with the numerical simulations that solve the Lindblad master equation and take pure dephasing into account, although the estimated pure dephasing rate is five times smaller than our previous experiments.
The experimental evaluation of the pure dephasing rate is an issue for future studies \cite{Abdumalikov2011, Lu2021}.

To demonstrate quantum-annealing applications using frequency chirping, we also conducted a measurement combining phase locking and correlated parametric oscillations, which can describe a general two-spin Ising model. 
We set the experimental parameters such that the bit-to-bit coupling and local fields favor the same-phase and opposite-phase states, respectively. 
We observed that the oscillation state with the highest observed probability depends on the relative magnitudes of bit-to-bit coupling and local fields. 
The results indicate quantum annealing in a two-KPO system with tunable Ising parameters using frequency-chirped drives.

Although we demonstrated the effectiveness of frequency chirping in KPO-based quantum annealing, there is room for parameter optimization, such as the oscillation amplitudes and the functional forms of the slopes of the drives. 
A scheme for the scalable embedding of all-to-all connected logical spins and the application of the scheme to KPO-based quantum annealing have been proposed \cite{Sourlas2005, Lechner2015, Puri2017a}. 
The scheme uses four-body interaction among physical spins to exclude physical-spin configurations inconsistent with the logical spins. 
Applying frequency-chirped drives in a multiple-KPO system using the scheme may enhance the effect of the four-body interaction and probability to obtain the oscillation states that can be decoded to logical Ising spins.

\begin{acknowledgments}
We thank Y. Kitagawa for his assistance in the device fabrication. 
We thank Y. Nakamura and Y. Urade for providing the JPAs used in this study.
The devices were fabricated in the Superconducting Quantum Circuit Fabrication Facility (Qufab) in National Institute of Advanced Industrial Science and Technology (AIST). 
This paper is based on results obtained from a project, JPNP16007, commissioned by the New Energy and Industrial Technology Development Organization (NEDO).

\end{acknowledgments}

\appendix

\section{Device-chip details}
\label{sup:device_detail}
\begin{figure}
	\includegraphics[width=\linewidth]{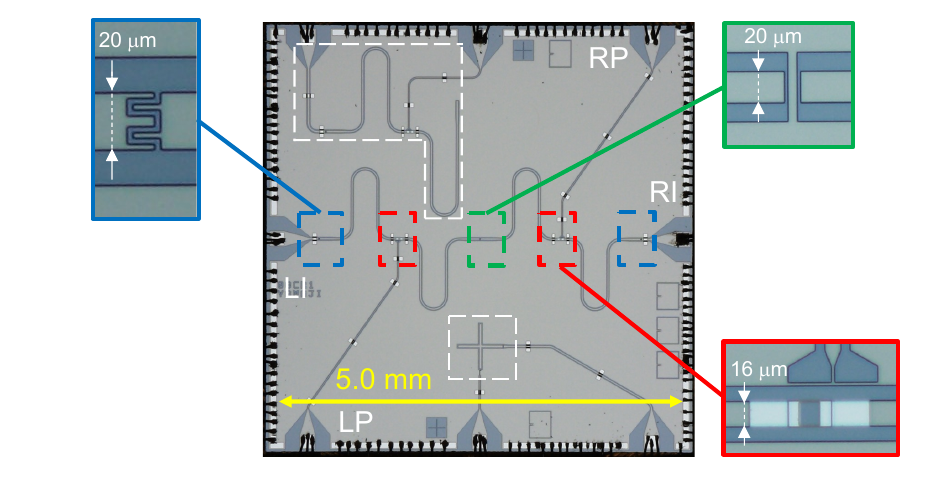}
	\caption{
Optical image of device chip studied in this paper. 
Red, blue, and green insets show magnified views around SQUIDs, I/O capacitors $C_{\rm in}$, and coupling capacitor $C_{\rm c}$, respectively, which are taken from the same type of chip in same lot. 
Uncoupled JPOs surrounded with white dashed lines are far detuned and not used in this study. 
Wire-bonding pads correspond to LI, LP, RI, and RP ports shown in Fig.~\ref{fig:sample}. 
}
	\label{fig:sample_pic}
\end{figure}

Figure~\ref{fig:sample_pic} shows the optical image of the device chip. 
The chip has four JPOs at the top-left, left, right, and bottom. 
In this study, we used two capacitively coupled JPOs on the left and right sides, the circuit diagram of which is shown in Fig.~\ref{fig:sample}, and the resonance frequencies of the other JPOs are far detuned. 
Each JPO has a half-wavelength (4.614~mm) resonator composed of coplanar waveguides (CPWs) with a designed characteristic impedance and phase velocity of 49.8~$\Omega$ and $0.398c$, respectively, where $c$ is the speed of light. 
A DC-SQUID (superconducting quantum interference device) with symmetric Josephson junctions (JJs) interrupts each resonator at its center. 
The resonators of the JPOs are capacitively coupled at the center of the chip with a coupling capacitance $C_{\rm c}$ of $0.6$~fF. 
The other sides of the resonators are connected to the LI and RI ports with an input/output (I/O) capacitance $C_{\rm in}$ of $3.4$~fF. 
The critical current of each JJ in the SQUIDs is estimated to be $0.83~{\rm \mu A}$ by fitting the DC-flux-bias dependence of the resonance frequencies of the JPOs \cite{Bourassa2012}. 
Each SQUID is inductively coupled to a pump line directly connected to the LP and RP ports. 
The CPWs on the chip have lithographically patterned airbridges to suppress parasitic slotline modes and reduce crosstalk between the JPOs.

We fabricated all the device components except JJs and airbridges using 100~nm-thick Nb film, which was sputtered on a high-resistive $380\ {\rm \mu m}$-thick Si substrate and patterned using dry etching with CF4 gas. 
The JJs were defined in a separate lithography step and fabricated by shadow evaporation after Ar-ion milling removed the Nb film's surface oxides. 
An additional photolithography using a spin-coated positive photoresist defined the contact pad of the airbridges. 
We deposited a 600~nm-thick sputtered Al film on the photoresist. 
We masked the bridge pattern with an additional positive photoresist and wet etched the Al layer except for the airbridges. 
Finally, we conducted ${\rm O}_2$ ashing and removed all the photoresists using an N-methyl-2-pyrrolidone(NMP)-based photoresist stripper.

\section{Operating point}
\label{sup:op.point}
The system Hamiltonian of the two-KPO system under the $\omega_{\rm p}$-resonance condition can be written using rotating-wave approximation in a frame rotating at the oscillation frequency $\omega_{\rm p}/2$ as follows \cite{Goto2018},
\begin{eqnarray}
\mathcal{H}_{\rm sys}/\hbar&=&(\mathcal{H}_{\rm \scalebox{0.5}{L}} +\mathcal{H}_{\rm \scalebox{0.5}{R}})/\hbar \nonumber \\
&&+g \left(e^{-{\rm i}\theta_{\rm p}/2}a_{\scalebox{0.5}{\rm L}}^\dag a_{\scalebox{0.5}{\rm R}} + e^{{\rm i}\theta_{\rm p}/2}a_{\scalebox{0.5}{\rm L}} a_{\scalebox{0.5}{\rm R}}^\dag \right), \label{eq:H_sys}\\
\mathcal{H}_j/\hbar&=&\frac{K_j}{2}a_j^{\dag 2} a_j^2 + \Delta_j a_j^\dag a_j + \frac{p_j}{2}\left(a_j^{\dag 2} + a_j^2\right) \nonumber \\
&&+ {\rm i}\Omega_{{\rm d}j} \left(e^{{\rm i}\theta_{{\rm s}j}} a_j^\dag - e^{-{\rm i}\theta_{{\rm s}j}} a_j \right), \label{eq:H_one}
\end{eqnarray}
where $\mathcal{H}_j$ is the one-JPO Hamiltonian, $K_j<0$ is the Kerr nonlinearity, $a_j$ is the annihilation operator of the fundamental mode, $p_j > 0$ is the amplitude of the two-photon drive (hereafter called pump amplitude) \cite{Yamaji2023}. 
The amplitude of the coherent states can be approximated to be $\alpha_j \simeq \sqrt{(p_j+\Delta_j)/|K_j|}$ by considering the one-photon drives, detuning, and coupling as perturbations \cite{Goto2018}. 
The eigenenergies of the simultaneous oscillation states are expressed as 
\begin{eqnarray}
E(s_{\rm \scalebox{0.5}{L}}, s_{\rm \scalebox{0.5}{R}})/\hbar &=& \sum_{j={\rm \scalebox{0.5}{L}}, {\rm \scalebox{0.5}{R}}} \left[\frac{K_j}{2}\alpha_j^4 + \Delta_j \alpha_j^2 + p_j \alpha_j^2 \right] \nonumber \\ 
&& - \left[-J_{\scalebox{0.5}{\rm LR}}s_{\rm \scalebox{0.5}{L}} s_{\rm \scalebox{0.5}{R}} + h_{\rm \scalebox{0.5}{L}}s_{\rm \scalebox{0.5}{L}} + h_{\rm \scalebox{0.5}{R}}s_{\rm \scalebox{0.5}{R}} \right]. \label{eq:Ising_appendix}
\end{eqnarray}
The second term is the Ising energy term, Eq.~(\ref{eq:Ising}), shown in the main text.

In the two-KPO experiments shown in the main text, the resonance frequencies of the JPOs are fixed to $\omega_{{\rm r}j}/2\pi \equiv\omega_{\rm r}/2\pi=9.88$~GHz.
The coupling between the two JPOs is estimated to be $g/2\pi=6.9$~MHz from the minimum frequency splitting $2g$ of the avoided level crossing at the $\omega_{\rm r}$-resonance condition, which is obtained from the transmission coefficient from the LI to RI ports. 
The external and internal photon-loss rates of the JPOs under the $\omega_{\rm r}$-resonance condition are estimated to be $\kappa_{\rm e\scalebox{0.5}{L(R)}}/2\pi=0.75(0.95)$~MHz and $\kappa_{\rm i\scalebox{0.5}{L(R)}}/2\pi=0.31(0.35)$~MHz, respectively, from the reflection coefficient with the other JPO far detuned. 
The Kerr nonlinearities are also evaluated to be $K_{\rm L(R)}/2\pi=-12.6$~MHz from the transition frequencies $\omega_{\rm r} + K/2$ of two-photon absorption between the vacuum state $\ket{0}$ and two-photon Fock state $\ket{2}$ \cite{Yamaji2022}. 
The transition frequency of the two-photon absorption is obtained from the probe-power dependence of the reflection coefficient with the other JPO far detuned. 
The JPOs are well in the single-photon Kerr regime, $\left|K_j\right|/\kappa_{{\rm a}j} \sim 10$, where $\kappa_{{\rm a}j}\equiv\kappa_{{\rm e}j} + \kappa_{{\rm i}j}$ is the total photon-loss rate. 

\begin{figure}
	\includegraphics[width=1.0 \linewidth]{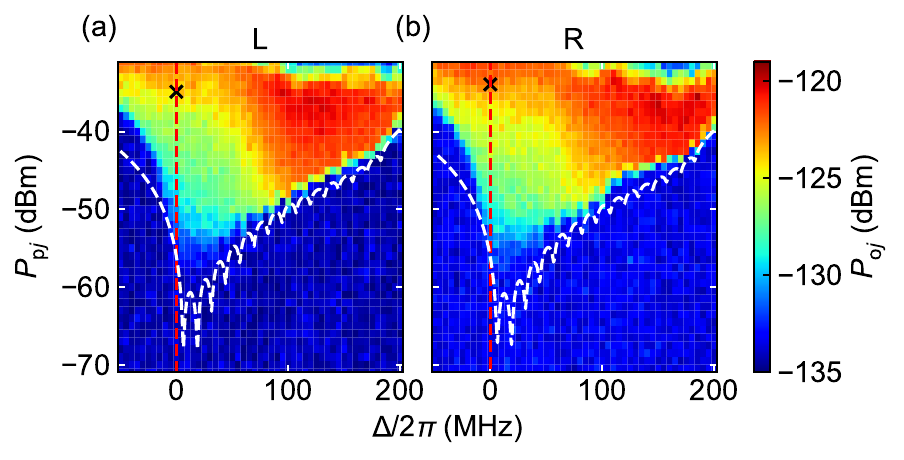}
	\caption{
CW parametric oscillations of KPOs L (a) and R (b). 
Resonance frequency, 9.88~GHz, is shown with red dashed line. 
Black crosses show operating point in two-KPO experiments presented in main text. White dashed lines show calculated pump power where mean photon number in KPOs is unity [$P_{{\rm o\scalebox{0.5}{L(R)}}}=-135(-134)\ {\rm dBm}$], which is close to noise floor of measurement. 
	}
	\label{fig:para_ex}
\end{figure}

Figure~\ref{fig:para_ex} shows the continuous-wave~(CW) parametric oscillations of the KPOs under the $\omega_{\rm r}$-resonance condition, where CW two-photon drives are individually applied to each KPO, not simultaneously to both KPOs, and the JPAs are turned off. 
The output power $P_{{\rm o}j}$ at $\omega_{\rm p}/2$ are measured using a spectrum analyzer as a function of the detuning $\Delta=\omega_{\rm r} - \omega_{\rm p}/2$ and pump power $P_{{\rm p}j}$. 
The amplitude of the two-photon drive $p_j$ is deduced from $P_{{\rm p}j}$ as follows \cite{Yamaguchi2024}, 
\begin{eqnarray}
p_j = \frac{1}{2}\left|\frac{d\omega_{{\rm r}j}}{di}\right|\sqrt{\frac{2}{Z_0 \times 1000}}10^{\frac{P_{{\rm p}j}}{20}},
\end{eqnarray} 
where $|d\omega_{{\rm r}\scalebox{0.5}{L(R)}}/di|=2\pi \times 2.60(2.65)~{\rm MHz/\mu A}$ is the first-order derivative of resonance frequencies with respect to the currents applied to the pump lines on the operating point, and $Z_0=50~{\Omega}$ is the characteristic impedance of the pump lines. 
The parametric oscillations are observed in the region where the output power is larger than the noise floor. 
The saturation power of the KPOs is $-120$~dBm. 
We compare the experiment with the analytical calculation of the steady state of the KPO \cite{Bartolo2016}, where we fixed all the parameters to the experimentally obtained values. 
The calculation reproduces the qualitative trend in the experiment. 
We observe the periodic structure of the oscillation threshold at $\Delta>0$ with an interval of $|K|$ both in the experiment and calculation, which was also reported in a previous study~\cite{Wang2019}. 

In the measurements presented in the main text, we set the pump power to be $P_{{\rm p\scalebox{0.5}{L(R)}}}=-35(-34)$~dBm [$p_{\rm \scalebox{0.5}{L(R)}}/2\pi=148(169)$~MHz (black crosses in Fig.~\ref{fig:para_ex}]), where the oscillation power is $P_{\rm o\scalebox{0.5}{L(R)}}=-124(-122)$~dBm, which are large enough for high signal-to-noise-ratio measurement and well below the saturation power of the JPOs.
The corresponding oscillation amplitudes are $\alpha^{\rm \scalebox{0.5}{CW}}_{\rm \scalebox{0.5}{L(R)}}=\sqrt{P_{\rm o\scalebox{0.5}{L(R)}} /(\hbar\omega_{\rm r} \kappa_{\rm e\scalebox{0.5}{L(R)}})}=3.4(4.0)$, which are consistent with the estimation obtained from the relation $\alpha_{\scalebox{0.5}{L(R)}}=\sqrt{p_{\scalebox{0.5}{L(R)}}/|K_{\scalebox{0.5}{L(R)}}|}=3.4(3.7)$ within the precision of 1 dB. 
We attribute the discrepancy to the uncertainties of the line attenuation and amplifier gain.

\section{Pulse sequence and parameters}
\label{sup:pulse_parameter}
We conducted the one- and two-KPO experiments in the main text using the pulse sequence shown in Fig.~\ref{fig:phaselock}(a) 
The trapezoidal two- and one-photon drives are applied simultaneously with the slope duration of $t_{\rm s}$ and turned off after the plateaus of $t_{\rm sp}$ and $t_{\rm pp}$, respectively, with the falling slope of $t_{\rm s}$. 
The slopes of the two-photon drive are nonlinear ($t^n$, $n>1$) to reduce nonadiabatic transitions at the start of parametric modulation \cite{Masuda2021}, although those of the one-photon drive are linear (see Appendix~\ref{sup:slope} for details). 
The output, which is induced by the two-photon drive, is amplified using amplifiers, including a Josephson parametric amplifier (JPA), and measured using the heterodyne measurement with an integration time of $t_{\rm r}$ (see Appendix~\ref{sup:meas-setup} for details). 
The heterodyne measurement is delayed from the start of the plateau by $t_{\rm rd}$ so as not to measure the one-photon drive reflected by the KPO, which is stronger (up to $-105$~dBm) than the oscillation power (-124 dBm). 
The interval between the one-photon drive and heterodyne measurement is $t_{\rm i}$. 
The detuning $\Delta/2\pi$ shifts linearly from $\Delta_0 \le 0$ to zero during the rising slopes of the trapezoidal pulses by synchronously chirping the frequencies of the two- and one-photon drives while maintaining the relationship that the two-photon-drive frequency is twice the one-photon-drive one.
The pulse sequence is repeated with a wait time of $t_{\rm w}$.

Table~\ref{tab:pulse} summarizes the pulse sequence parameters of the measurements in the main text.
The pump slope $t_{\rm pp}$ of the two-KPO experiments is set to maximize correlation between simultaneous parametric oscillations (see Appendix~\ref{sup:slope_delay} for details). 
Table~\ref{tab:param} summarizes the other parameters of the experiments shown in Figs.~\ref{fig:phaselock}-\ref{fig:boltzmann}.
The histograms, Figs.~\ref{fig:phaselock}(b) and \ref{fig:correlation}(a), are obtained by integrating $10^4$ pulses.
Each data in Figs.~\ref{fig:phaselock}(c), \ref{fig:correlation}(b, c), and \ref{fig:boltzmann} is obtained using $10^4$ pulses.

\begin{table*}
\centering
\begin{tabular}{lccccccccc} \hline
measurement & $t_{\rm s}$ & $t_{\rm sp}$ & $t_{\rm pp}$ & $t_{\rm r}$ & $t_{\rm rd}$ & $t_{\rm i}$ & $t_{\rm w}$ & $n$  \\
            & [ns] & [ns] & [ns] & [ns] & [ns] & [ns] & [ns] &   \\ \hline
One-KPO, phase locking (Figs.~\ref{fig:phaselock} and \ref{fig:1bit_simulation})      & 100 & 100 & 800  & 400 & 300 & 100 & 920 & 5.0 \\ \hline
Two-KPO, w/o phase locking (Fig.~\ref{fig:correlation})   & 400 &     & 600  & 400 & 100 &     & 520 & 2.5 \\ \hline
Two-KPO, w/ phase locking (Fig.~\ref{fig:boltzmann})   & 400 & 100 & 1100 & 400 & 600 & 100 & 660 & 2.5 \\ \hline
\end{tabular}
\caption{
Pulse sequence parameters. 
Parameters in the main text, one-KPO experiment (Figs.~\ref{fig:phaselock} and \ref{fig:1bit_simulation}), two-KPO experiment without phase locking (Fig.~\ref{fig:correlation}), and two-KPO experiment with phase locking (Fig.~\ref{fig:boltzmann}), are shown. 
} 
\label{tab:pulse}
\end{table*}

\begin{table}
\centering
\resizebox{\columnwidth}{!}{
\begin{tabular}{rcccccccc} \hline
 & $P_{\rm s\scalebox{0.5}{L}}$ & $P_{\rm s\scalebox{0.5}{R}}$ & $\theta_{\rm s\scalebox{0.5}{L}}$ & $\theta_{\rm s\scalebox{0.5}{R}}$ & $\theta_{\rm p}$ & $\Delta_0/2\pi$ & $N_{\rm pulse}$ \\
 & [dBm] & [dBm] & [rad] & [rad] & [rad] & [MHz] & \\ \hline
Fig.~\ref{fig:phaselock}(b)    & -132  &       & $1.1\pi$ && & -20  & $10^4$ \\
(c) & swept & & swept & && -20 & $10^4$/data \\ \hline
Fig.~\ref{fig:correlation}(a)   &       &       & & & $0.0\pi$ & -20 & $10^4$ \\
(b) &&&&& swept & -20 & $10^4$/data \\
(c) &&&&& swept & swept & $10^4$/data \\ \hline
Fig.~\ref{fig:boltzmann}     & swept & swept & $1.5\pi $ & $0.5\pi$ & $0.0\pi$ & -20 & $10^4$/data \\ \hline
\end{tabular}
}
\caption{
Parameters of experiments shown in Figs.~\ref{fig:phaselock}-\ref{fig:boltzmann}.
} 
\label{tab:param}
\end{table}

\section{Measurement setup}
\label{sup:meas-setup}
 \begin{figure}
	\includegraphics[width=\linewidth]{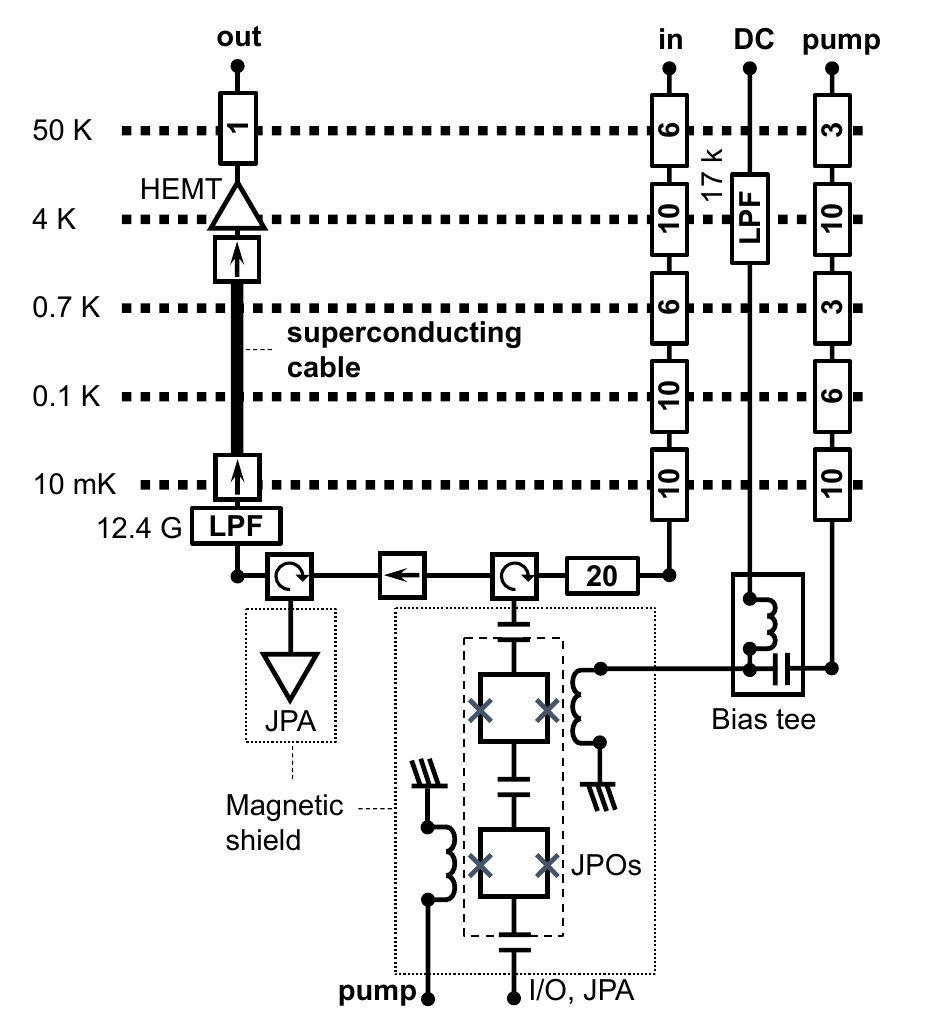}
	\caption{
Circuit diagram in dilution refrigerator. 
Horizontal dotted lines represent temperature stages of refrigerator. 
Diagram shows I/O line, pump line, DC line, and JPA for only one of two JPOs. 
The DC and pump lines for JPA are also not shown. 
Rectangles on lines represent cryogenic attenuators and low-pass filters (LPFs), where values inside and beside rectangles show attenuation in decibels and cut-off frequencies, respectively. 
Squares with circular and straight arrows represent circulators and isolators, respectively. 
	}
	\label{fig:rf_fridge}
\end{figure}

Figure~\ref{fig:rf_fridge} shows the circuit diagram in the dilution refrigerator.
The diagram is almost the same as in our previous experiments \cite{Yamaji2022, Yamaji2023}, except that the JPAs amplify the outputs of the JPOs. 
The JPAs are impedance-matched Josephson parametric amplifiers (IMPAs) \cite{Mutus2014, Roy2015, Urade2021}. 
The resonance frequencies of the IMPAs are tuned to $\sim10$~GHz then parametrically modulated with a pump frequency of 20~GHz. 
The IMPAs amplify input signals by three-photon mixing, and the gain at 9.88 GHz is 13(16)~dB for JPO L(R), respectively. 

The input, output, pump, and DC lines are connected to instruments in a room-temperature environment, where radio-frequency switches Keysight L8990M select the CW and time-domain (TD) measurement systems. 
The TD measurement system comprises a calibration-free microwave-pulse generation system called Quel-1 \cite{Sumida2024, Kurimoto2025_arXiv}. 
The one-photon drives at 9.88~GHz are generated by mixing 11.4-GHz CW microwaves and 20-MHz intermediate frequency (IF) trapezoidal pulses with 1.5-GHz digital upconversion using digital-to-analog converters. 
Arbitrary waveform generators make the trapezoidal pulses with a sampling rate of 500~MHz, where frequency chirping is executed by varying the IF frequency. 
The two-photon drives are generated using a doubler and another pulse-generation system with the same configuration as the one-photon drives. 
The output is down-converted using the 11.4-GHz CW microwaves and recorded using analog-to-digital converters with a sampling rate of 500~MHz and 1.5-GHz digital downconversion. 
The {\it IQ} amplitudes are then deduced using the Fourier transformation at 20~MHz.

\section{Numerical method}
\label{sup:numerical}

We simulated the dynamics of the KPOs using the same method as in a previous study~\cite{Yamaji2023}. The master equations describing the one- and two-KPO experiments are as follows: 
\begin{eqnarray}
\frac{d\rho_{1j}}{dt} &=& - \frac{i}{\hbar} [\mathcal{H}_j,\rho_{1j}] \nonumber\\ 
&+& 
\frac{\kappa_{{\rm a}j}^\ast}{2} \mathcal{D}[a_j, \rho_{1j}]
+ \gamma_j \mathcal{D}[a_j^\dagger a_j,\rho_{1j}], \\
\frac{d\rho_2}{dt} &=& - \frac{i}{\hbar} [\mathcal{H}_{\rm sys},\rho_2] \nonumber\\ 
&+& \sum_{j={\rm \scalebox{0.5}{L},\scalebox{0.5}{R}}} \Big{[} \frac{\kappa_{{\rm a}j}^\ast}{2} \mathcal{D}[a_j, \rho_2]
+ \gamma_j \mathcal{D}[a_j^\dagger a_j,\rho_2] \Big{]},
\end{eqnarray} 
where $\rho_{1j}$ and $\rho_2$ are the density matrices of the one- and two-KPO systems, respectively, $\mathcal{D}[\hat{O}, \rho] \equiv 2\hat{O}\rho \hat{O}^\dagger - \hat{O}^\dagger \hat{O} \rho - \rho \hat{O}^\dagger \hat{O}$, and $\kappa_{{\rm a}j}^\ast$ is the total photon-loss rate without contribution from pure dephasing. The spectroscopically measured internal photon-loss rates $\kappa_{{\rm i}{\rm \scalebox{0.5}{L(R)}}}/2\pi=0.31 (0.35)$~MHz include contribution from pure dephasing. 
The internal and total photon-loss rates without contribution from pure dephasing are $\kappa_{{\rm i}j}^\ast=\kappa_{{\rm i}j}-2\gamma_j$ and $\kappa_{{\rm a}j}^\ast=\kappa_{{\rm e}j}+\kappa_{{\rm i}j}^\ast$, respectively.

The occurrence probabilities of the oscillation states are deduced from the Husimi Q functions as 
\begin{eqnarray}
\xi_{1j\pm}(t) &=& \frac{1}{\pi} \int\limits_{
\substack{
{\rm sign} ({\rm Re}[\alpha_j])\\
=\pm1
}
}
d\alpha_j \bra{\alpha_j}\rho_{1j}(t)\ket{\alpha_j},\\
\xi_{2\pm}(t)&=&\frac{1}{\pi^2} \iint\limits_{
\substack{
{\rm sign}({\rm Re}[\alpha_{\rm \scalebox{0.4}{L}}]{\rm Re}[\alpha_{\rm \scalebox{0.4}{R}}])\\
=\pm 1
}
}
d\alpha_{\rm \scalebox{0.5}{L}}d\alpha_{\rm \scalebox{0.5}{R}} \bra{\alpha_{\rm \scalebox{0.5}{L}}}\bra{\alpha_{\rm \scalebox{0.5}{R}}}\rho_2(t)\ket{\alpha_{\rm \scalebox{0.5}{L}}}\ket{\alpha_{\rm \scalebox{0.5}{R}}}, \nonumber \\
\end{eqnarray} 
where $\xi_{1j\pm}(t)$ is the occurrence probabilities of the oscillation states $\ket{\pm \alpha_j}$, $\xi_{2\pm}(t)$ is the occurrence probabilities of the same-phase and opposite-phase simultaneous parametric oscillations, and $\alpha_j$ is regarded as a complex integral variable in these equations. These probabilities are averaged during the period corresponding to the integration time of the heterodyne measurement.

\section{Dependence on drive slopes}
\label{sup:slope}
\begin{figure}
	\includegraphics[width=\linewidth]{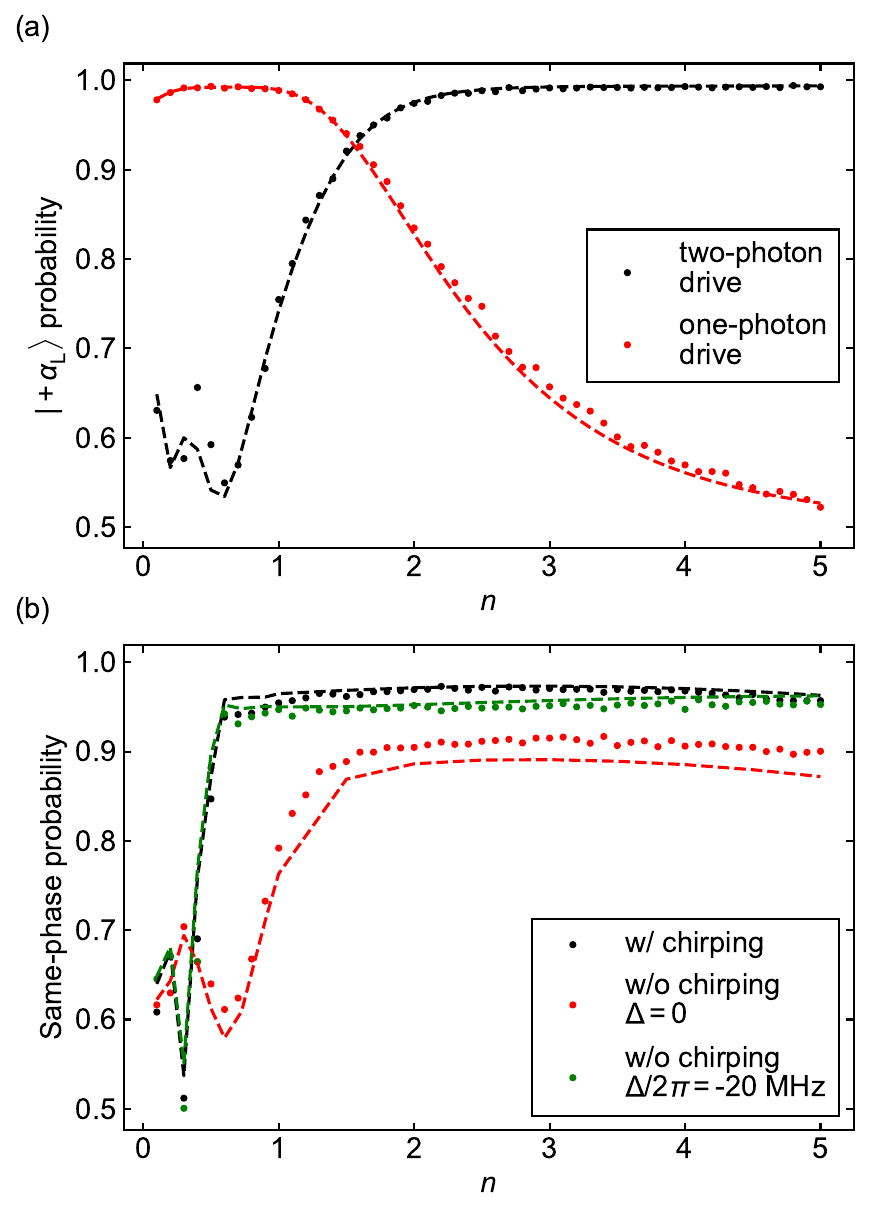}
	\caption{
Dependence on exponent of drive slopes. 
(a) Phase locking of KPO L. 
Exponent $n$ of slope $t^n$ of two- and one-photon drives are swept under same condition as data with frequency chirping in Fig.~\ref{fig:phaselock}(d), where we fix $P_{\rm s \scalebox{0.5}{L}}=-105$~dBm and $\theta_{\rm s\scalebox{0.5}{L}}=1.5\pi$. 
Black and red circles show experimental data sweeping exponent of two- and one-photon drives, while other exponent is fixed to $n=1$ (one-photon drive) and $n=2.5$ (two-photon drive). 
Dashed lines show simulated exponent dependence assuming $\gamma_{\rm \scalebox{0.5}{L}}/2\pi=6.8~{\rm kHz}$. 
(b) Correlation of simultaneous parametric oscillations with and without frequency chirping. 
Exponent of two-photon-drive slope is swept under the same condition as the data at $\theta_{\rm p}=0.0\pi$ shown in Fig.~\ref{fig:correlation}(b). 
Black circles show experimental data obtained with frequency chirping. 
Red and green circles show experimental data obtained without frequency chirping at $\Delta=0$ and $\Delta/2\pi=-20$~MHz, respectively. 
Dashed lines show simulated exponent dependence assuming $\gamma_{\rm \scalebox{0.5}{L(R)}}/2\pi=7.7~{\rm kHz}$. 
Each data point in (a-b) is obtained using $10^4$ pulses. 
	}
	\label{fig:xn_dependence}
\end{figure}

The time evolution of the two-KPO system during parametric modulation depends on the functional form of the drive slopes. We show how we determined the functional forms used in the one- and two-KPO experiments in the main text. We set the functional form as $t^n$ in the one- and two-KPO experiments, fixed other experimental parameters, and swept the exponent $n$ to explore the optimal functional form that maximizes the locking probabilities and correlation between simultaneous parametric oscillations. Figure~\ref{fig:xn_dependence} shows the dependence on the slope exponent of phase locking and correlation of simultaneous parametric oscillations. As shown in Fig.~\ref{fig:xn_dependence}(a), the phase-locking probability is an increasing function of the exponent of the two-photon drive because the large exponent reduces nonadiabatic transitions from the vacuum state at the start of the parametric oscillations \cite{Masuda2021}. On the contrary, the phase-locking probability is a decreasing function of the exponent of the one-photon drive, possibly due to the nonadiabatic transitions close to the plateau of the one-photon drives. The locking probability saturates at the exponent of $n=5$ (two-photon drive) and $n=1$ (one-photon drive). The experimental data are consistent with the simulation solving the master equation. It is worth noting that the exponent dependence is sensitive to the oscillation amplitude, and the consistency between the experiment and simulation limits the uncertainty of the oscillation amplitude below 0.5~dB. 
The correlation of simultaneous parametric oscillations also depends on the slope exponent of the two-photon drives, as shown in Fig.~\ref{fig:xn_dependence}(b). 
The correlation with frequency chirping has its maximum at around $n=2.5$. 
The simulation involving $\gamma_{\rm \scalebox{0.5}{L(R)}}/2\pi=7.7$~kHz qualitatively reproduces the experimental data with and without frequency chirping.
On the basis of the slope-exponent dependence, we fix the slope exponent of the two- and one-photon drives to be $n=5(2.5)$ and $n=1(1)$ for the one-KPO (two-KPO) experiments in the main text, respectively.

\section{Slope and delay dependence of correlation}
\label{sup:slope_delay}
\begin{figure}[b]
	\includegraphics[width=\linewidth]{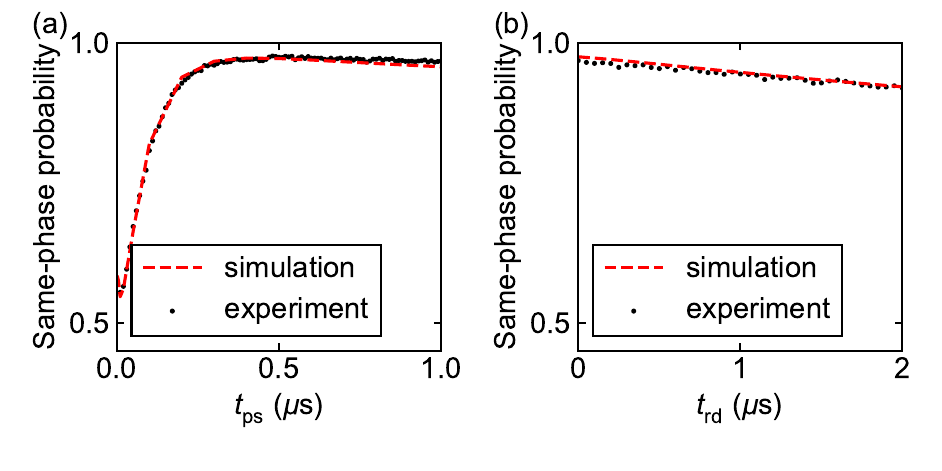}
	\caption{
Slope and readout-delay dependence of ferromagnetic correlation at $\theta_{\rm p}=0.0\pi$ with frequency chirping. 
Same pulse sequence as that in Fig.~\ref{fig:correlation}, except $t_{\rm s}$, $t_{\rm rd}$, and $t_{\rm pp}$, is used. 
(a) Slope duration $t_{\rm s}$ dependence of same-phase probability, where $t_{\rm rd}=100$~ns and $t_{\rm pp}=600$~ns. 
(b) Readout delay $t_{\rm rd}$ dependence of same-phase probability, where $t_{\rm s}=400$~ns and $t_{\rm pp}=2500$~ns. 
Black circles show experimental data obtained from $10^4$ pulses. Red dashed lines show simulated dependence assuming $\gamma_{\rm \scalebox{0.5}{L(R)}}/2\pi=7.7$~kHz. 
	}
	\label{fig:sd_dependence}
\end{figure}

Since the time evolution of the two-KPO system also depends on the duration of the  drive slope $t_{\rm s}$, we measured the $t_{\rm s}$ dependence of the correlation between simultaneous parametric oscillations to determine $t_{\rm s}$ in the two-KPO experiment presented in the main text. 
Figure~\ref{fig:sd_dependence}(a) shows the $t_{\rm s}$ dependence of the same-phase probability at $\theta_{\rm p}=0.0\pi$. 
The $t_{\rm s}$ dependence is an increasing function in the range $t_{\rm s} < 400~{\rm ns}$ since the adiabatic transition from the initial vacuum state to the same-phase states with the lowest Ising energy requires the slow change in the system Hamiltonian to satisfy the adiabatic condition. 
The correlation saturates at $t_{\rm s}=400\sim600~{\rm ns}$, and the $t_{\rm s}$ dependence becomes a decreasing function in the range $t_{\rm s} > 600~{\rm ns}$ since pure dephasing causes the phase flipping of parametric oscillations, as shown in the readout delay $t_{\rm rd}$ dependence [Fig.~\ref{fig:sd_dependence}(b)]. 
The correlation is a monotonically decreasing function of $t_{\rm rd}$, where the flipping rate is estimated to be 27~kHz by fitting the $t_{\rm rd}$ dependence with an exponential function. 
The simulations involving $\gamma_{\rm \scalebox{0.5}{L(R)}}/2\pi=7.7$~kHz are quantitatively consistent with the experimental data. 
On the Basis of the $t_{\rm s}$ and $t_{\rm rd}$ dependencies, we fixed $t_{\rm s}=400$~ns and minimized $t_{\rm rd}$ in the two-KPO experiments in the main text.

\bibliography{master}

\end{document}